\documentclass{jfm}

\usepackage{color}
\usepackage{amssymb,overpic}
\usepackage{amsmath}
\usepackage{graphicx}
\usepackage{lipsum}
\usepackage{multirow}

\definecolor{blue}{rgb}{0, 0.4470, 0.7410}
\definecolor{red}{rgb}{0.8500, 0.1250, 0.0480}
\definecolor{green}{rgb}{0.4660, 0.6740, 0.1880}

\graphicspath{.}

\begin{document}

\shorttitle{Resolvent analysis on the origin of two-dimensional transonic buffet} 
\shortauthor{Kojima et al.} 

\title{Resolvent analysis on the origin of \\ two-dimensional transonic buffet}         

\author
 {
  Yoimi Kojima\aff{1}
  \corresp{\email{y-kojima@st.go.tuat.ac.jp}},
  Chi-An Yeh\aff{2},
  Kunihiko Taira\aff{2}
  \\
  \and
  Masaharu Kameda\aff{1}
  }

\affiliation
{
\aff{1}
Department of Mechanical Systems Engineering, Tokyo University of Agriculture and Technology, Koganei, Tokyo, 184-8588, Japan
\aff{2}
Department of Mechanical and Aerospace Engineering, University of California, Los Angeles, CA 90095, USA
}

\maketitle

\begin{abstract}
Resolvent analysis is performed to identify the origin of two-dimensional transonic buffet over an airfoil.
The base flow for the resolvent analysis is the time-averaged flow over a NACA 0012 airfoil at a chord-based Reynolds number of $2\,000$ and a free-stream Mach number of $0.85$. 
We reveal that the mechanism of buffet is buried underneath the global low-Reynolds number flow physics.  At this low-Reynolds number, the dominant flow feature is the von K\'arm\'an shedding.  However, we show that with the appropriate forcing input, buffet can appear even at a Reynolds number that is much lower than what is traditionally associated with transonic buffet.  The source of buffet is identified to be at the shock foot from the windowed resolvent analysis,
which is validated by companion simulations using sustained forcing inputs based on resolvent modes.  
We also comment on the role of perturbations in the vicinity of the trailing edge.  The present study not only provides insights on the origin of buffet but also serves a building block for low-Reynolds number compressible aerodynamics in light of the growing interests in Martian flights.
\end{abstract}

\section{Introduction}
\label{sec:intro}

Transonic flow over an airfoil exhibits complex dynamics from the 
interplay among the unsteady shock motion, flow separation due to the shock/boundary-layer interaction, and the generation and propagation of pressure waves.
Under certain range of angles of attack and Mach number, the shock wave oscillates vigorously creating a highly unsteady flow \citep{lee2001self}. This shock wave oscillation phenomenon known as transonic buffet is caused by a self-sustaining aerodynamic instability. 
The large-scale motion of the shock induces violent fluctuations in aerodynamic forces and leads to structural vibrations.
The level of unsteadiness induced by transonic buffet can be so large that it can compromise the integrity of aircraft structure and flight safety.
For these reasons, transonic buffet is a major limiting factor for aircraft flight envelope and has been a focus of many studies, as summarized recently by \citet{giannelis2017review}.

Various experimental and numerical studies have been devoted to the characterization of transonic buffet. Past studies have revealed that typical oscillating frequency of shock waves is very low with the chord-based Strouhal number being approximately 0.06 \citep{deck2005numerical, jacquin2009experimental}. This low frequency is distinguished from other unsteady phenomena around the airfoil related to the velocity fluctuations in the separation region and vortex shedding in the wake \citep{sartor2014stability}.  More recently, some researchers have reported that a transonic buffet on a three-dimensional wing has span-wise instability with a much higher frequency where $0.2 < St < 0.6$ \citep{dandois2016experimental, Ohmichi:AIAAJ2018}.

Separation on the airfoil induced by shock-wave/boundary layer interaction (SWBLI) plays an important role in transonic buffet.  Early experiments \citep{seegmiller1978, LevyJr.1978} showed that the onset of shock oscillation occurred when the separation generated at the shock foot reaches the trailing edge of the airfoil. Recent numerical studies \citep{Iovnovich2012, Grossi2014, fukushima2018wall} reported that unsteadiness of the separation is closely tied to buffet.  When the shock moves downstream from its upstream position, the shock becomes weaker and the flow behind the shock remains attached. In contrast, when the shock moves upstream from its downstream location, the shock becomes stronger, and causes large-scale separation.

A number of studies have analyzed the self-sustained mechanism of shock wave oscillation.
\citet{lee2001self} argued that the pressure wave propagating between the shock wave and the trailing edge of the airfoil plays a critical role feedback to maintain shock oscillation. More recently, global linear stability analysis performed by \citet{crouch2009origin} extracted the growth rate of the buffet fluctuation with respect to the time-averaged flow field based on the eigenvalue problem of linearized Navier-Stokes operator.  
They examined the linear stability of transonic flow around a two-dimensional NACA 0012 airfoil obtained from Reynolds-averaged Navier--Stokes (RANS) simulations. Analyzing the critical angle of attack and Mach number of the buffet onset, the critical values reasonably agreed with the experimental results by \citet{mcdevitt1985static}. 
Nonetheless, the origin of the transonic buffet is still under debate, with open questions on validity of the linear stability formalism for base flows that are not the exact solution of the Navier--Stokes equations.

Recently, \citet{sartor2014stability} employed global stability, adjoint, and resolvent analysis to examine transonic buffet flow around a NACA0012 airfoil at $Re = 3 \times 10^6$. 
At the buffet frequency, they used resolvent analysis to find that the essential origin of shock unsteadiness is at the shock foot region on the suction side of the airfoil. 
There are also reports including \citet{Nitzsche2009} that identified various sources that can trigger buffet such as the motion of flap, the pitch oscillation, and the translation of the airfoil. 
Based on these observations, various methods has been developed to control buffet, including vortex generators \citep{mccormick1993shock} and trailing-edge deflection techniques \citep{gao2017active}.

The objective of the present study is to systematically determine the origin of two-dimensional transonic buffet flow around a NACA0012 airfoil at a low-Reynolds number of $Re = 2\,000$ in the absence of any modeling terms. 
We reveal the input-output relationship for transonic buffet
with the resolvent analysis, which can extract not only the frequency response but also identify the forcing (input) and response (output) modes of flow systems \citep{Trefethen1993, jovanovic2005componentwise, mckeon2010critical, Yeh2019, taira2017modal, taira2019modal}. 
To validate that the identified source of buffet is indeed the driving mechanism for sustained oscillations, we perform direct numerical simulations (DNS) to show that buffet can appear even in low Reynolds number flows under appropriate conditions.
This analysis not only provides the fundamental insights into transonic buffet but also serves as a building block for low-Reynolds number compressible aerodynamics ($Re \approx O(10^3 \sim 10^5)$).  This area of aerodynamics has traditionally been overlooked but is now becoming important for developing high-efficiency wings and propulsion systems for unmanned aircraft on Mars \citep{Anyoji2015, Munday:JA15, Koning2019}. 

Below, we discuss the problem description, simulation approach, and resolvent analysis in \S \ref{sec:setup}.  The insights gained from resolvent analysis are offered in \S \ref{sec:results} with discussions on the source of buffet.  Our findings are validated by forcing the flow at the identified source to stimulate the emergence of buffet.  At last, we offer concluding remarks in \S \ref{sec:conclusion}.

\section{Problem setup}
\label{sec:setup}

\subsection{Problem description}

We consider two-dimensional laminar transonic flows over a  NACA 0012 airfoil at a free-stream Mach number $M_\infty \equiv U_\infty / a_\infty = 0.85$, a chord-based Reynolds numbers of $Re_{L_c} \equiv U_\infty L_c / \nu_\infty = 2~000$, and an angle of attack of $\alpha = 3^\circ$. Here, $a_\infty$ is the free-stream sonic speed, $L_c$ is the chord length, and $\nu_\infty$ is the free-stream kinetic viscosity. Prandtl number is set to $Pr = 0.7$. While not reported, we also considered $\alpha = 1^\circ$ and $5^\circ$. 
As we obtained analogous results for these angles of attack, we only present the representative case of $\alpha = 3^\circ$ herein.  
Shock waves form around the airfoil under these conditions, making this base flow an appropriate candidate for this study.

\subsection{Flow simulation}

\begin{figure}
    \centering
    \includegraphics[width=.90\textwidth]{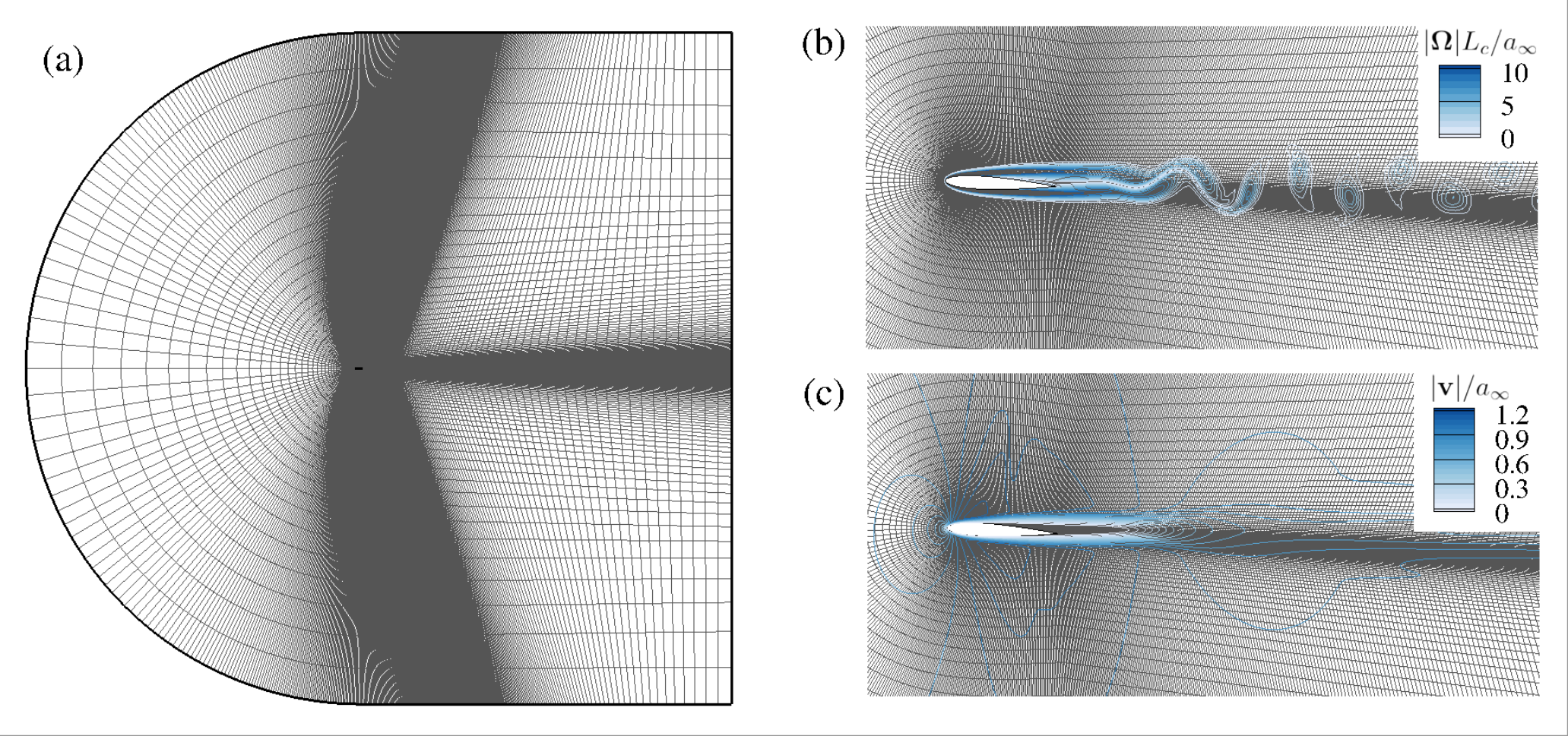}
    \caption{
    The computational grid used for numerical simulation and resolvent analysis. (a) The whole computational domain. The enlarged views around the airfoil are shown with the (b) instantaneous magnitude of normalized vorticity and (c) time-averaged velocity fields.
    }
    \label{figMesh}
\end{figure}

We simulate transonic flows over the airfoil by solving the compressible Navier--Stokes equations in a two-dimensional setting using the finite-volume solver {\it{CharLES}}, which is second-order accurate in space and third-order accurate in time \citep{bres2017unstructured}.  We employ a second-order fully unstructured essentially non-oscillatory (ENO) method \citep{Shi:JCP2002} for shock capturing.  A C-shaped hexahedral mesh is used for the simulations, as shown in figure \ref{figMesh}.  The computational domain has an extent of $x_c/L_c \in [-50, 50]$ and $y_c/L_c \in [-50, 50]$, where $x_c$ and $y_c$ represent the chord-wise and chord-normal directions, respectively. The airfoil is positioned with its leading edge at the origin.  The domain is discretized into $70~000$ cells with 100 nodes on each side of the airfoil, 90 nodes along the wake and 80 nodes in the normal direction where the minimum wall-adjacent $\Delta y$ is set to be $\Delta y / L_c = 1.42 \times 10^{-3}$.
A grid convergence study has been conducted with respect to the time-averaged flow to ensure sufficient spatial resolution for this computational grid. 
At the far-field boundary, the free-stream condition is prescribed as $[\rho, v_{x_c}, v_{y_c}, T] = [\rho_\infty, U_\infty \cos \alpha, U_\infty \sin \alpha, T_\infty]$, where $\rho$ is density and $T$ is temperature.  The free stream angle is changed through the prescription of the far-field flow velocities $v_{x_c}$ and $v_{y_c}$ in $x_c$ and $y_c$ directions, respectively.  No-slip adiabatic condition is prescribed over the airfoil.  Along the outlet boundary, a sponge layer \citep{Freund:AIAAJ97} is applied over $x_c/L_c \in [40, 50]$ with the target state being the running-averaged flow over $t U_\infty/L_c = 1$.  The time integration is performed at a constant Courant--Friedrichs--Lewy number of $1$.

\subsection{Resolvent analysis}
\label{secMethodResolvent}

The objective of the present study is to reveal the input--output relationship for two-dimensional transonic buffet and identify the origin of the energetic shock-wave oscillations.  In this study, we focus our resolvent analysis to two-dimensional modes which are essential in all buffet phenomena \citep{Iovnovich2012, sartor2014stability, Ohmichi:AIAAJ2018}. Note that resolvent analysis itself is not limited to the two-dimensional flow. Further resolvent analysis of transonic buffet over the three-dimensional wings could extract the spanwise modes, which have been reported in previous studies \citep{Iovnovich:AIAAJ2015, Ohmichi:AIAAJ2018}.

We consider the Reynolds decomposition of the flow variable $\boldsymbol{q}(\boldsymbol{x}, t) = [\rho, v_x, v_y, T]$ into the sum of time-averaged base state $\bar{\boldsymbol{q}}(\boldsymbol{x})$ and the statistically stationary fluctuating component $\boldsymbol{q}'(\boldsymbol{x}, t)$, which enables us to express the Navier--Stokes equation as
\begin{equation}
    \partial_t \boldsymbol{q}' = \mathsfbi{L}_{\bar{\boldsymbol{q}}} \boldsymbol{q}'+ \mathsfbi{B} \boldsymbol{f}',
    \label{eq:re-dec-final}
\end{equation}
where $\mathsfbi{L}_{\bar{\boldsymbol{q}}}$ is the linearized Navier--Stokes operator about the mean flow $\bar{\boldsymbol{q}}$ \citep{mckeon2010critical} and $\mathsfbi{B}$ is the input matrix that can spatially window the forcing input $\boldsymbol{f}'$.  
We construct the discrete linear operator $\mathsfbi{L}_{\bar{\boldsymbol{q}}}$, incorporating the boundary conditions of $[\rho', v_{x_c}', v_{y_c}', \nabla_n T'] = 0$ over the airfoil and the far field, and $\nabla_n [\rho', v_{x_c}', v_{y_c}', T'] = 0$, at the computational outlet, where $\nabla_n$ denotes the surface-normal gradient. The spacial discretization for $\mathsfbi{L}_{\bar{\boldsymbol{q}}}$ is performed on the same mesh used in the flow simulation, as shown in figure \ref{figMesh}.
When the time-averaged flow is chosen as the base state in resolvent analysis, the finite-amplitude nonlinear terms with respect to $\boldsymbol{q}'$ are incorporated in $\boldsymbol{f}'$, which can be interpreted as a sustained forcing input within the natural feedback system \citep{mckeon2010critical}.  Moreover, we consider an observable output vector $\boldsymbol{y}$ given by 
$
    \boldsymbol{y} = \mathsfbi{C} \boldsymbol{q}',
$
such that spatial windowing $\mathsfbi{C}$ can be applied in general \citep{Jeun:PoF2016,Schmidt:JFM2018}. 

With the Fourier representation $\boldsymbol{q}'(\boldsymbol{x}, t) = 
	\int_{-\infty}^{\infty}
	    \hat{\boldsymbol{q}}_{\omega}(\boldsymbol{x}) 
    e^{- i\omega t} {\rm d}\omega$, 
equation (\ref{eq:re-dec-final}) can be expressed in frequency space as 
\begin{equation}
    -i\omega \hat{\boldsymbol{q}}_\omega = \mathsfbi{L}_{\bar{\boldsymbol{q}}} \hat{\boldsymbol{q}}_\omega 
    + \mathsfbi{B} \hat{\boldsymbol{f}}_\omega
    \quad
    \text{with}
    \quad
    \hat{\boldsymbol{y}}_\omega 
    = \mathsfbi{C} 
    \hat{\boldsymbol{q}}_\omega.
    \label{eq:NS_FreqDomain}
\end{equation}
The input--output relationship between the input (forcing) $\hat{\boldsymbol{f}}_\omega$ and the output (response) $\hat{\boldsymbol{y}}_\omega$ at a specified frequency $\omega$ becomes 
\begin{equation}
	\hat{\boldsymbol{y}}_\omega 
	= \mathsfbi{R}_{\bar{\boldsymbol{q}}}(\omega)
	\hat{\boldsymbol{f}}_\omega,
\quad
\text{where}
\quad
    \mathsfbi{R}_{\bar{\boldsymbol{q}}}(\omega) 
    = 
    \mathsfbi{C}
    \left[ -i\omega \mathsfbi{I} - \mathsfbi{L}_{\bar{\boldsymbol{q}}} \right]^{-1} 
    \mathsfbi{B}.
    \label{eq:resol-op}
\end{equation}
Here, $\mathsfbi{R}_{\bar{\boldsymbol{q}}}(\omega)$ is referred to as the resolvent (operator).  It serves as a transfer function that amplifies (or attenuates) the harmonic forcing input $\hat{\boldsymbol{f}}_\omega$ and maps it to the response $\hat{\boldsymbol{y}}_\omega$.  In this study, the base flows are found to be stable according the eigenvalues of the linear operators $\mathsfbi{L}_{\bar{\boldsymbol{q}}}$.  Therefore, we consider real-valued frequency $\omega$ in constructing $\mathsfbi{R}_{\bar{\boldsymbol{q}}}(\omega)$ for our resolvent analysis \citep{jovanovic2004modeling, Yeh2019}.

Resolvent analysis identifies the dominant directions along which $\hat{\boldsymbol{f}}_\omega$ can be most amplified through $\mathsfbi{R}_{\bar{\boldsymbol{q}}}(\omega)$ to form the corresponding response $\hat{\boldsymbol{y}}_\omega$.  This is accomplished by performing the singular value decomposition (SVD) of the resolvent
\begin{equation}
    \mathsfbi{R}_{\bar{\boldsymbol{q}}}(\omega) = \mathsfbi{Y}\boldsymbol{\Sigma}\mathsfbi{F}^*,
    \label{eq:svd}
\end{equation}
where $\mathsfbi{F}^*$ denotes the Hermitian of $\mathsfbi{F}$. 
Resolvent analysis interprets left and right singular vectors $\mathsfbi{Y} = [\hat{\boldsymbol{y}}_1, \hat{\boldsymbol{y}}_2, \dots, \hat{\boldsymbol{y}}_m]$ and $\mathsfbi{F} = [\hat{\boldsymbol{f}}_1, \hat{\boldsymbol{f}}_2, \dots, \hat{\boldsymbol{f}}_m]$, respectively, as response modes and forcing modes, with the magnitude-ranked singular values $\boldsymbol{\Sigma} = {\text{diag}}(\sigma_1, \sigma_2, \dots, \sigma_m) $ being the amplification (gain) for the corresponding forcing--response pair at frequency $\omega$.  We examine the modal structures of the dominant forcing and response modes $[\hat{\boldsymbol{f}}_1, \hat{\boldsymbol{y}}_1]$ to study the origin of transonic buffet over this canonical airfoil.

The application of spatial window to the global response $\hat{\boldsymbol{q}}_\omega$ can be implemented by specifying $\mathsfbi{C}$ as a diagonal weight matrix with unit weights inside the window and zeros otherwise.  This window is chosen to highlight the location over the suction surface where oscillatory motion of the shock wave appears and limits the output to be within inside of window, as is illustrated in figure \ref{figAveFlow}.  While matrix $\mathsfbi{B}$ for forcing can be designed in a similar manner, we use $\mathsfbi{B} = \mathsfbi{I}$ such that there is no restriction on the region of forcing to identify the source of buffet.  The use of window $\mathsfbi{C}$ in the resolvent analysis can help reveal the optimal energy amplification from forcing to a local response in the shock region over the suction surface of the airfoil.

\section{Results}
\label{sec:results}

\subsection{Base flow}
\label{secRsltBase}

\begin{figure}
\centering
    \begin{tabular}{ccccc}
        \begin{overpic}[width=0.45\textwidth]{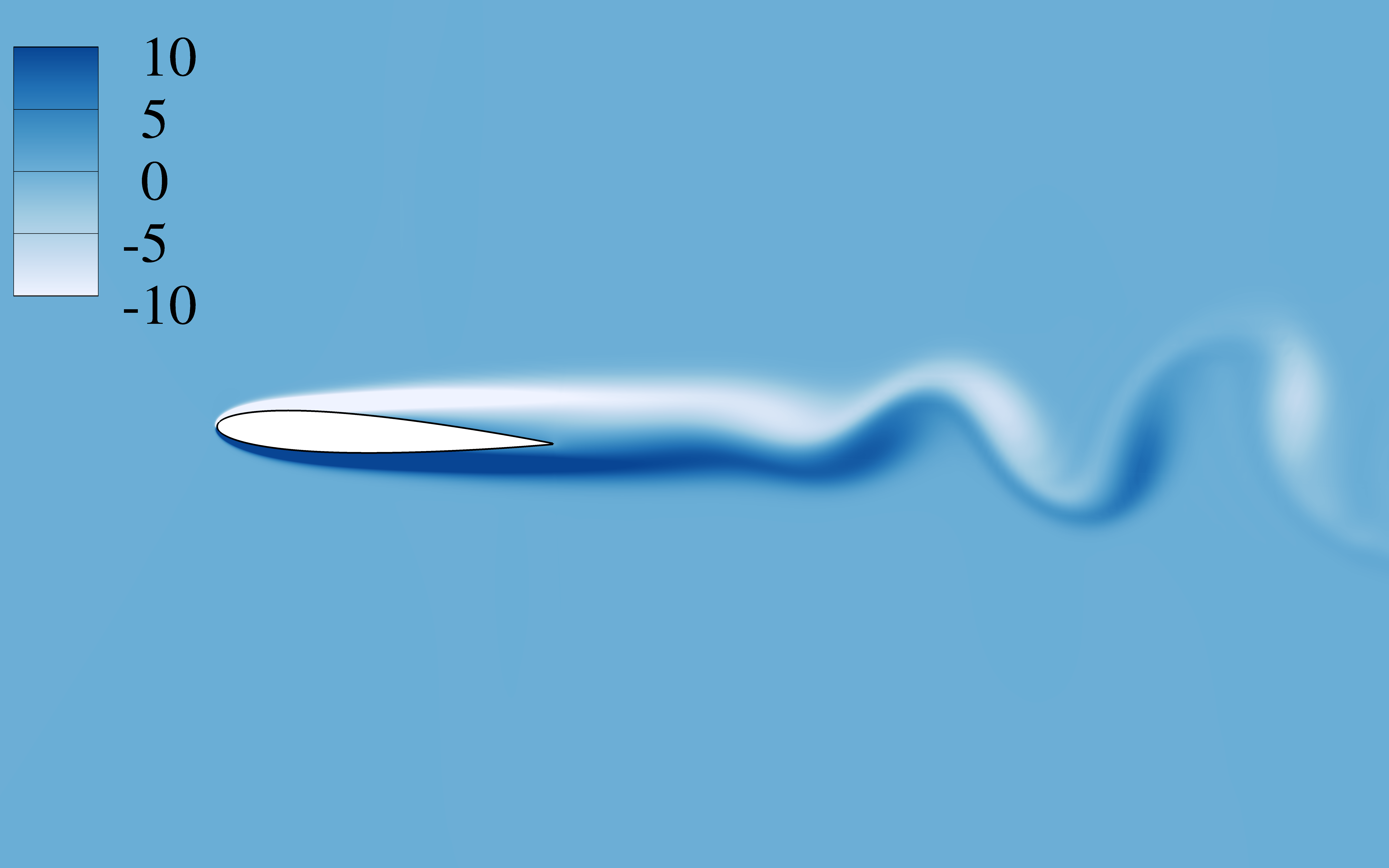}
    	    \put (-9, 57) {\indexsize (a)}
	    \end{overpic}
	    &
	    &&&
	    \begin{overpic}[width=0.36\textwidth]{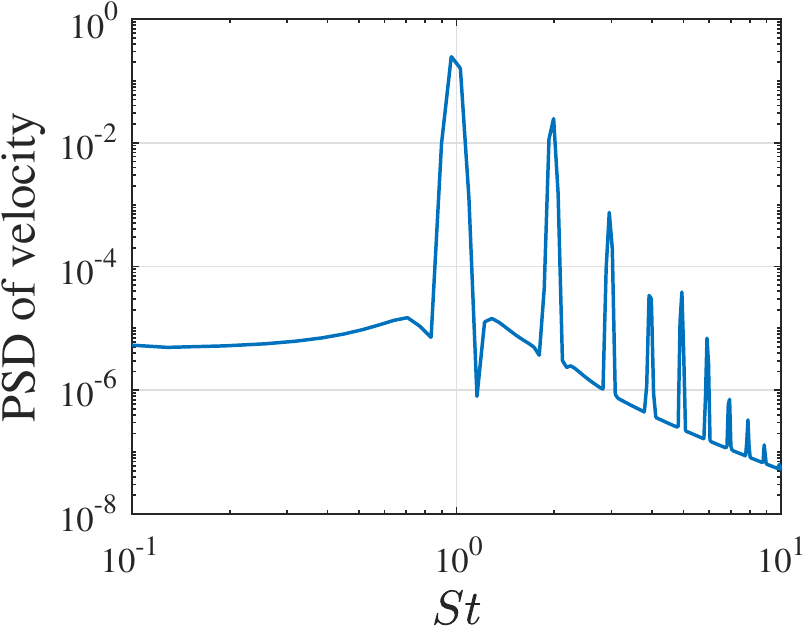}
    	    \put (-7, 72) {\indexsize (b)}
	    \end{overpic}
    \end{tabular}
    \caption{Unsteady transonic flow over a NACA0012 airfoil.  (a) Instantaneous vorticity field ($ \Omega_z L_c / a_\infty$).  (b) Power spectral density of $v_y/a_\infty$ probed at $(x, y)/L_c = (2.10, -0.11)$.
    }
    \label{figInstVel}
\end{figure}

We simulate the unsteady two-dimensional flow over a NACA 0012 airfoil at $\alpha = 3^\circ$ for $M_\infty = 0.85$ and $Re_{L_c} = 2\,000$. An instantaneous vorticity field from DNS is shown in figure \ref{figInstVel}(a). 
The laminar vortex sheets generated from the pressure and suction sides of the wing rolls into vortices in the wake.  This representative transonic flow at low Reynolds number exhibits a distinct shedding frequency of $St \equiv f L_c/U_\infty = 1.0$.  This frequency can be detected from the power spectral density (PSD) of the velocity probe at $(x, y)/L_c = (2.10, -0.11)$, along with its harmonics, in figure \ref{figInstVel}(b).  For this flow, we do not directly observe buffet over the airfoil. \citet{Nitzsche2009} investigated the frequency response of shock waves to small perturbations and found that the external forcing to transonic flow around an airfoil can trigger shock wave oscillations even if the base flow is stable. He has reported that the maximum gain was observed close to the buffet onset angle. The present resolvent analysis holds great potential to reveal the optimal input-output relationship with respect to the onset of buffet.

\begin{figure}
    \centering
    \begin{tabular}{cccc}
        \begin{overpic}[height=0.27\textwidth]{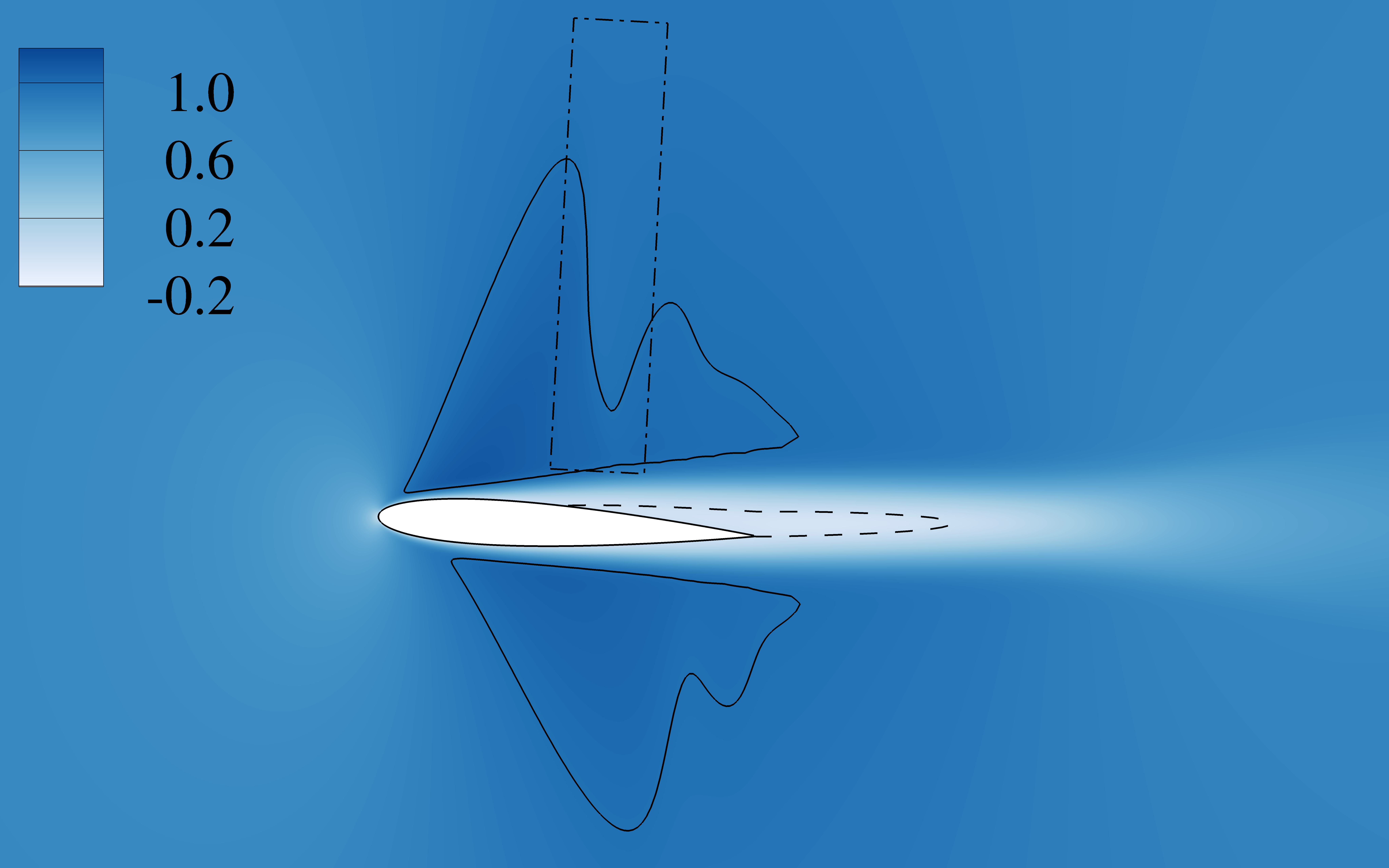}
    	    \put (-8, 57) {\indexsize (a)}
	    \end{overpic}
	    &
	    &&
	    \begin{overpic}[height=0.27\textwidth]{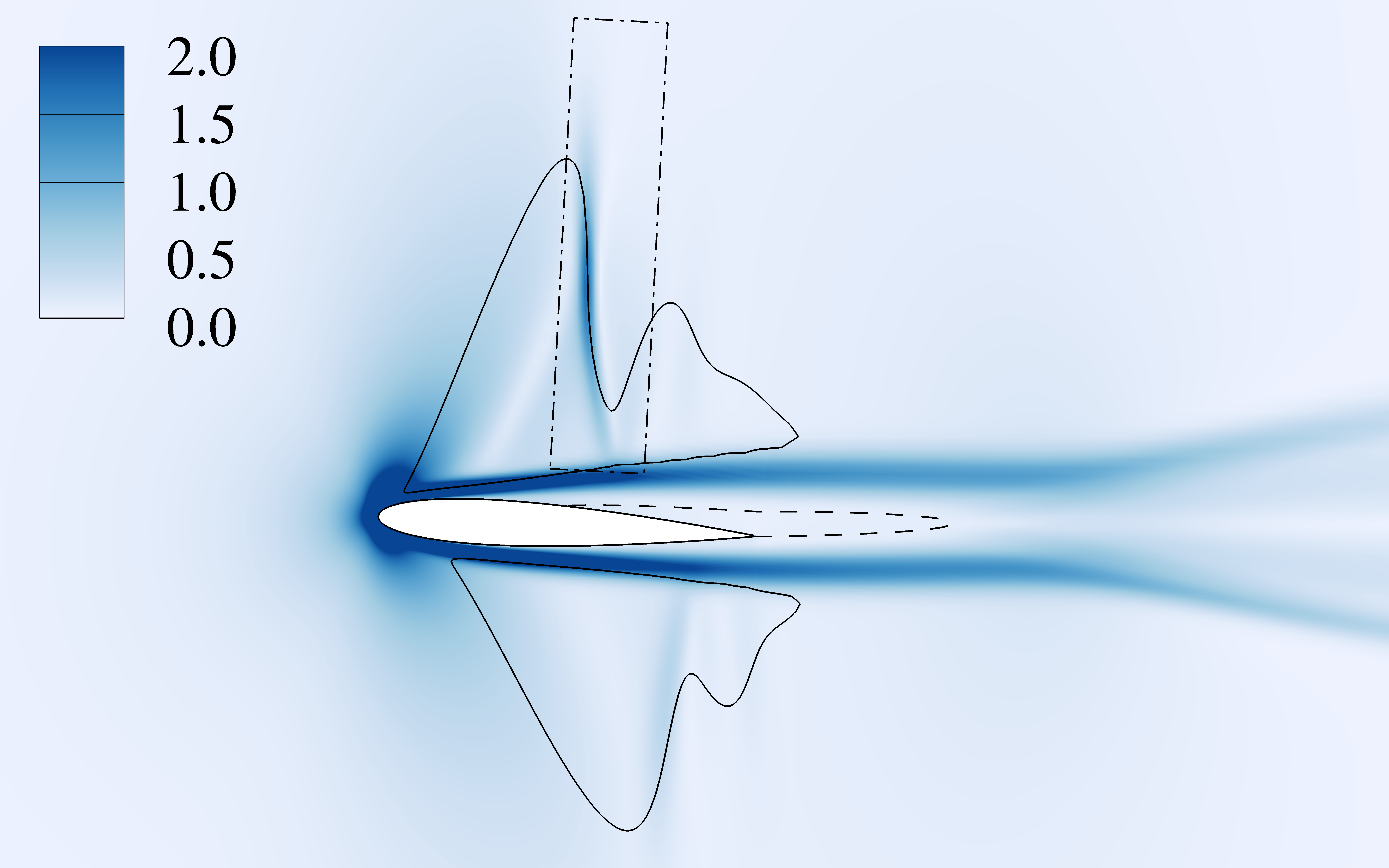}
    	    \put (-8, 57) {\indexsize (b)}
	    \end{overpic}
    \end{tabular}
    \caption{
    Time-averaged flow field around the NACA0012 airfoil. (a) Streamwise velocity ($v_x / a_\infty$). (b) Numerical schlieren with density gradient magnitude ($\|\nabla \rho\| L_c / \rho_\infty$).
    Regions enclosed by the solid and broken curves represent the supersonic and recirculation regions, respectively. The dashed-dotted box depicts the spatial window for the filtered resolvent analysis.
    }
    \label{figAveFlow}
\end{figure}

To perform resolvent analysis, we consider the time-averaged flow to be the base state.  The temporal averaging is performed over $t U_\infty/L_c \in [0, 350]$ to ensure statistical convergence.  This long window for average allows us to set the minimum frequency in the resolvent analysis at $2.9\times10^{-3}$, which is sufficiently small compared with the dominant buffet frequency of $St \approx 0.07$ \citep{jacquin2009experimental,sartor2014stability}.
The time-averaged stream-wise velocity and the numerical shlieren fields are shown in figure \ref{figAveFlow}. 
The solid and broken curves in figures \ref{figAveFlow}(a) and (b) indicate the contour lines for $M = 1$ and $v_x = 0$, respectively.  
The area enclosed in the solid curve corresponding to the supersonic region, while the region surrounded by the broken curves represents the separated flow region.  
Weak shock waves are observed in the supersonic regions.  The supersonic region on the suction side is further divided into two stages of acceleration--deceleration phases according to the $\lambda$-type structure of the shock waves \citep{delery1985shock}. 
The first stage corresponds to the initial flow acceleration near the leading edge.  The flow surpasses the sonic speed and decelerates after the first shock wave.  Over the airfoil, the flow undergoes the second stage of acceleration and forms the second shock structure near the trailing edge.  Strong inverse pressure gradient due to $\lambda$-type shock waves are observed aft of the first acceleration.

\subsection{Resolvent analysis}
\label{secRsltResolvent}

\begin{figure}
    \centering
    \begin{tabular}{ccccc}
        \begin{overpic}[height=0.338\textwidth]{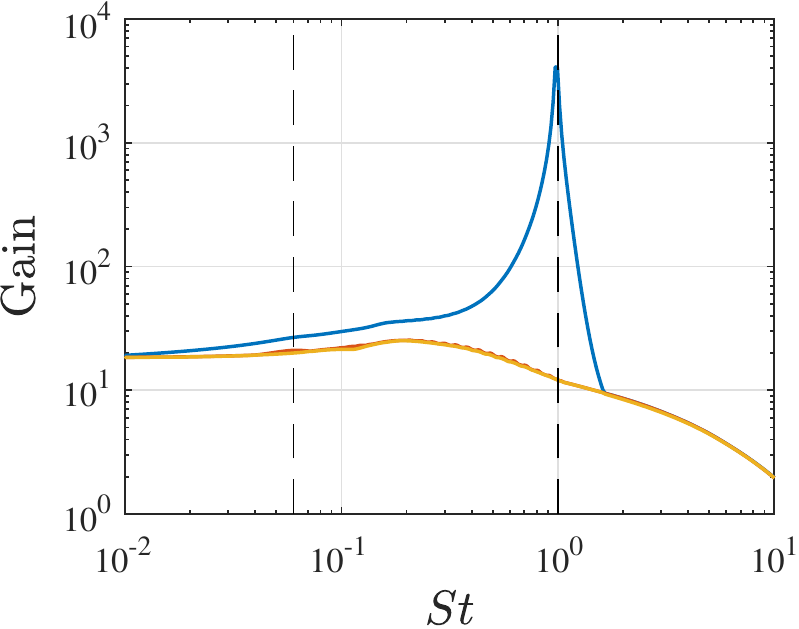}
    	    \put (-8, 73) {\indexsize (a)}
	    \end{overpic}
	    &
	    &&&
	    \begin{overpic}[height=0.326\textwidth]{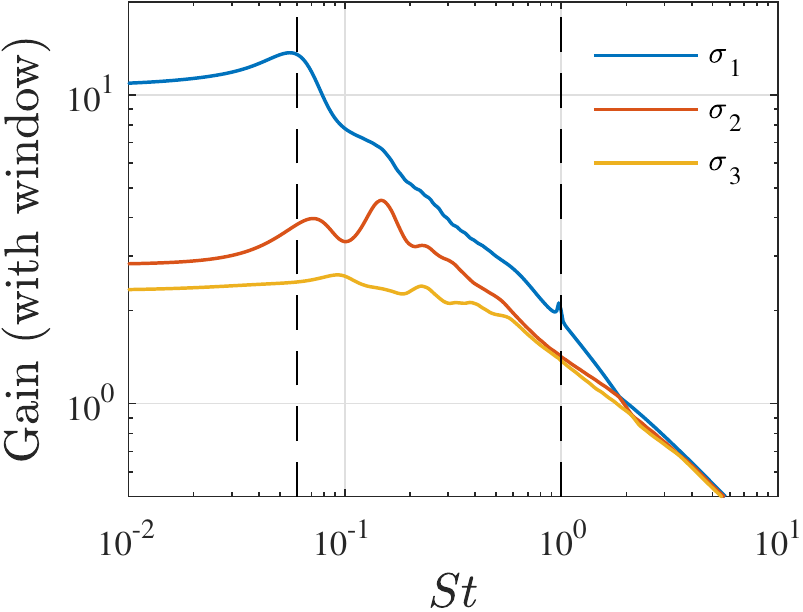}
    	    \put (-8, 74) {\indexsize (b)}
	    \end{overpic}
    \end{tabular}
    
    \caption{(a) Original and (b) windowed resolvent gains over $St$.}
    \label{figSvDist}
\end{figure}

As transonic buffet is characterized by the energetic shock-wave oscillation over the airfoil, the cause of buffet may be uncovered by closely studying the input--output process at the relevant fluctuation frequency using the resolvent analysis described in Section \ref{secMethodResolvent}.  
Let us first present in figure \ref{figSvDist}(a) the distribution of the leading gain $\sigma_1$ over a range of frequencies from resolvent analysis without windowing.  The leading gain $\sigma_1$ shows a distinct peak at $St = 1$ corresponding to the von K\'arm\'an shedding frequency.  However, there is no other distinct peaks appearing in the gain distribution for $\sigma_1$.  This is expected since the dominant physics for this low Reynolds number base flow is the wake shedding.  We do not observe noticeable peaks even for subdominant modes.   

\begin{figure}
    \centering
    \begin{tabular}{cccc} \hline
          \multicolumn{2}{c}{Response mode} & \multicolumn{2}{c}{Forcing mode} \\ 
          $v_x$ component & $v_y$ component & $v_x$ component & $v_y$ component \\ \hline \includegraphics[width=0.235\textwidth]{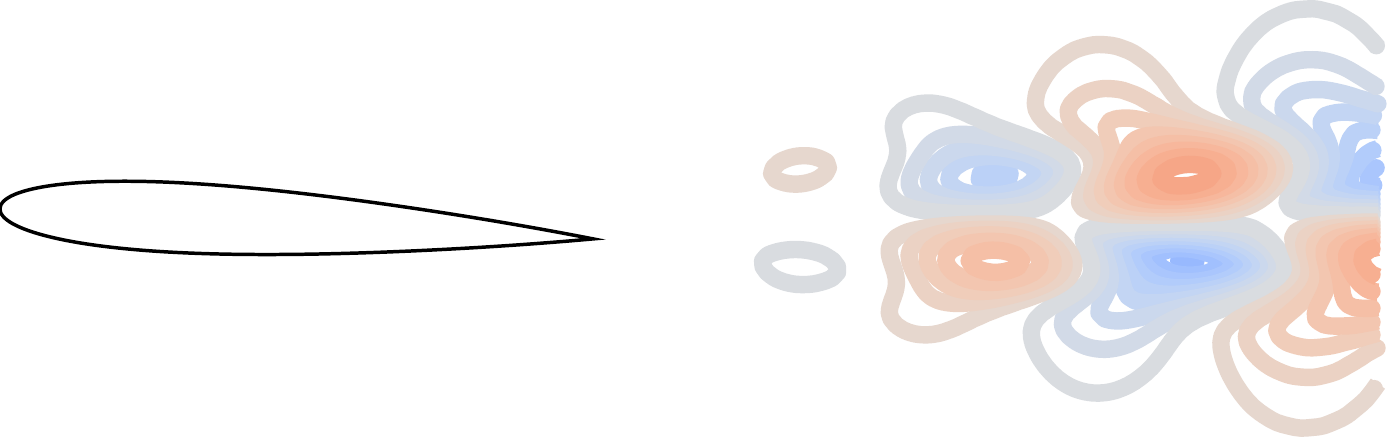} & \includegraphics[width=0.235\textwidth]{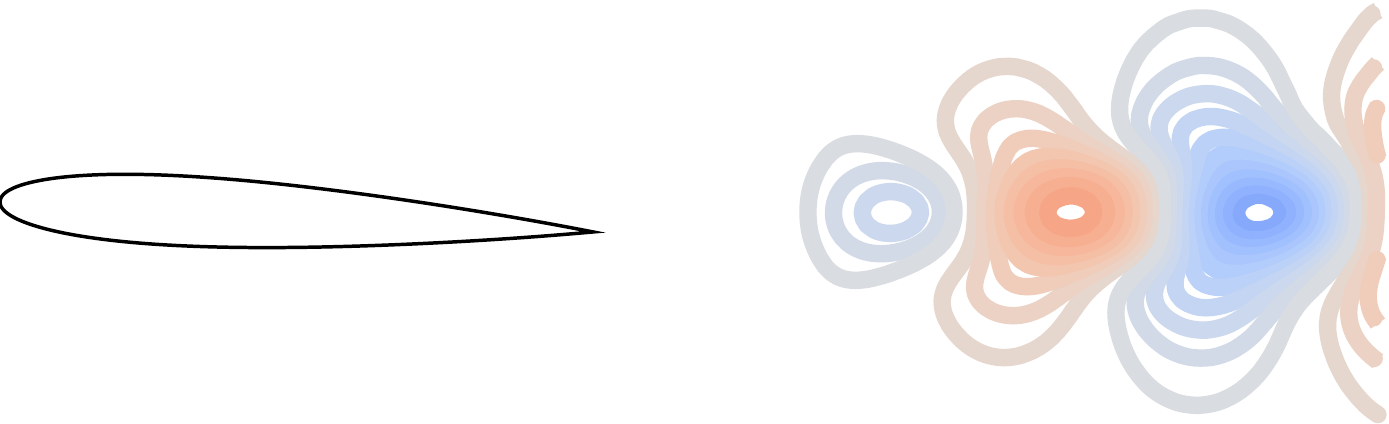} & \includegraphics[width=0.235\textwidth]{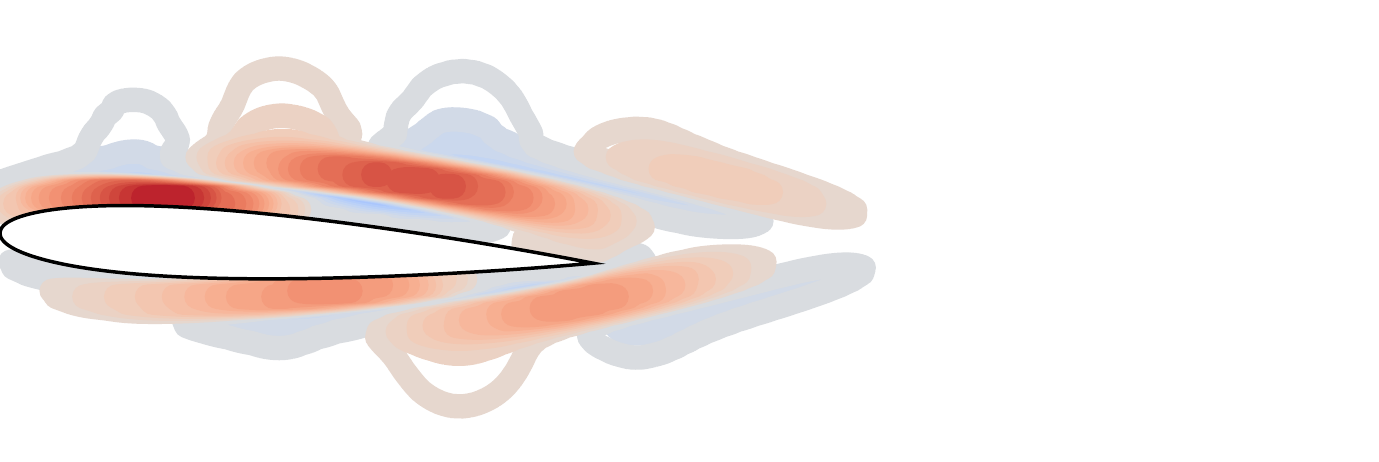} & \includegraphics[width=0.235\textwidth]{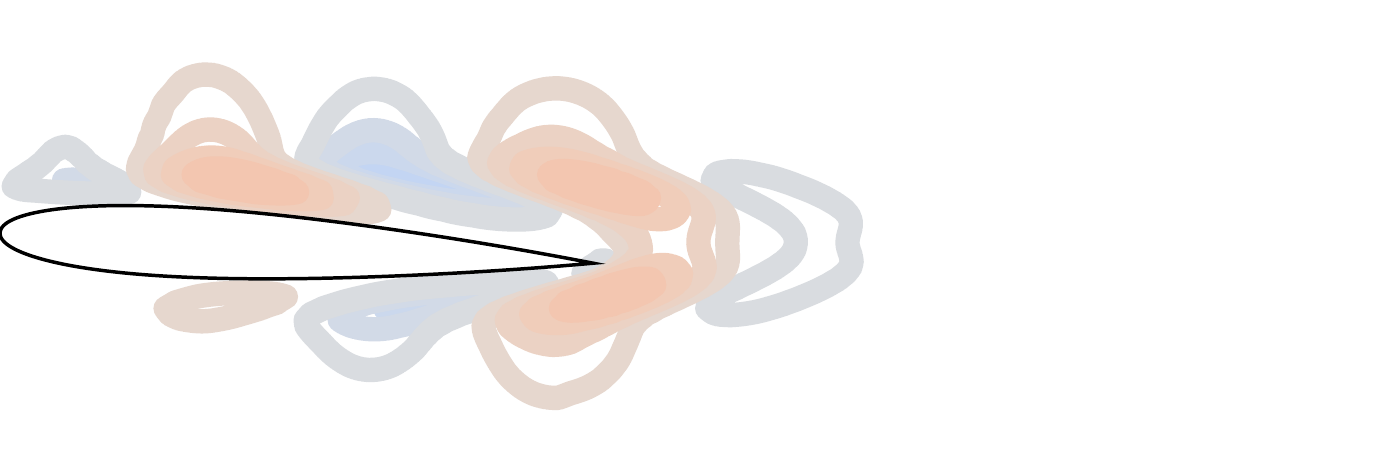} \\ \hline
    \end{tabular}
    \caption{Response and forcing modes normalized by the velocity magnitude for $St=1$.
    }
    \label{figMode_St1}
\end{figure}

Let us present the dominant resolvent modes for $St = 1$ without windowing.  Since the gain peaks at the shedding frequency, response modes exhibit the modal shapes that are representative of convective dynamics in the wake, as visualized in figure \ref{figMode_St1}.  The forcing modes are present in the vicinity of the airfoil where shear is large, as it can be observed from figure \ref{figAveFlow}.  This is in line with our understanding of the formation of wake vortices from the roll up of shear layers.  These observations are also in agreement with modal analysis for subsonic flows over airfoils \citep{Zhang:PF16, Yeh2019}.

Next, let us consider the gain distribution from the windowed resolvent analysis.  Here, the output is windowed by operator $\mathsfbi{C}$ with a spatial profile that is restricted to the region in the vicinity of the shock, as shown in figure \ref{figAveFlow}.  With the output of resolvent analysis focusing on the fluctuations close to the standing shock, we find that the dominant gain $\sigma_1$ shows a peak at the buffet frequency of $St = 0.06$, which is in agreement with past numerical and experimental findings \citep{deck2005numerical, jacquin2009experimental, dandois2016experimental, sartor2014stability}, which were conducted at high Reynolds numbers of $Re_{L_c} > O(10^6)$.
For subdominant gains, we observe peaks appearing at superharmonics that capture smaller-scale oscillations.  We can observe a minor peak in $\sigma_1$ at $St = 1$.  However, the overall amplification process for the shock region is focused toward the low-frequency components at the buffet frequency.  This suggests that the amplification physics for buffet is buried underneath the global input-output dynamics at this low Reynolds number.  We also note that the conclusive assessment on the buffet physics is robust against the choice of the response window.  A prominent peak at the buffet frequency $St = 0.06$ with similar forcing structure is still revealed using a different window that covers the entire suction surface over $x_c \in [0, 1]$ and $y_c \in [0, \infty)$.

\begin{figure}
    \centering
    \begin{tabular}{ccccc} \hline
           & \multicolumn{2}{c}{Response modes} & \multicolumn{2}{c}{Forcing modes} \\
           & $v_x$ components & $v_y$ components & $v_x$ components & $v_y$ components \\ \hline
          \rotatebox{90}{~~~~~Original} & \includegraphics[width=0.22\textwidth]{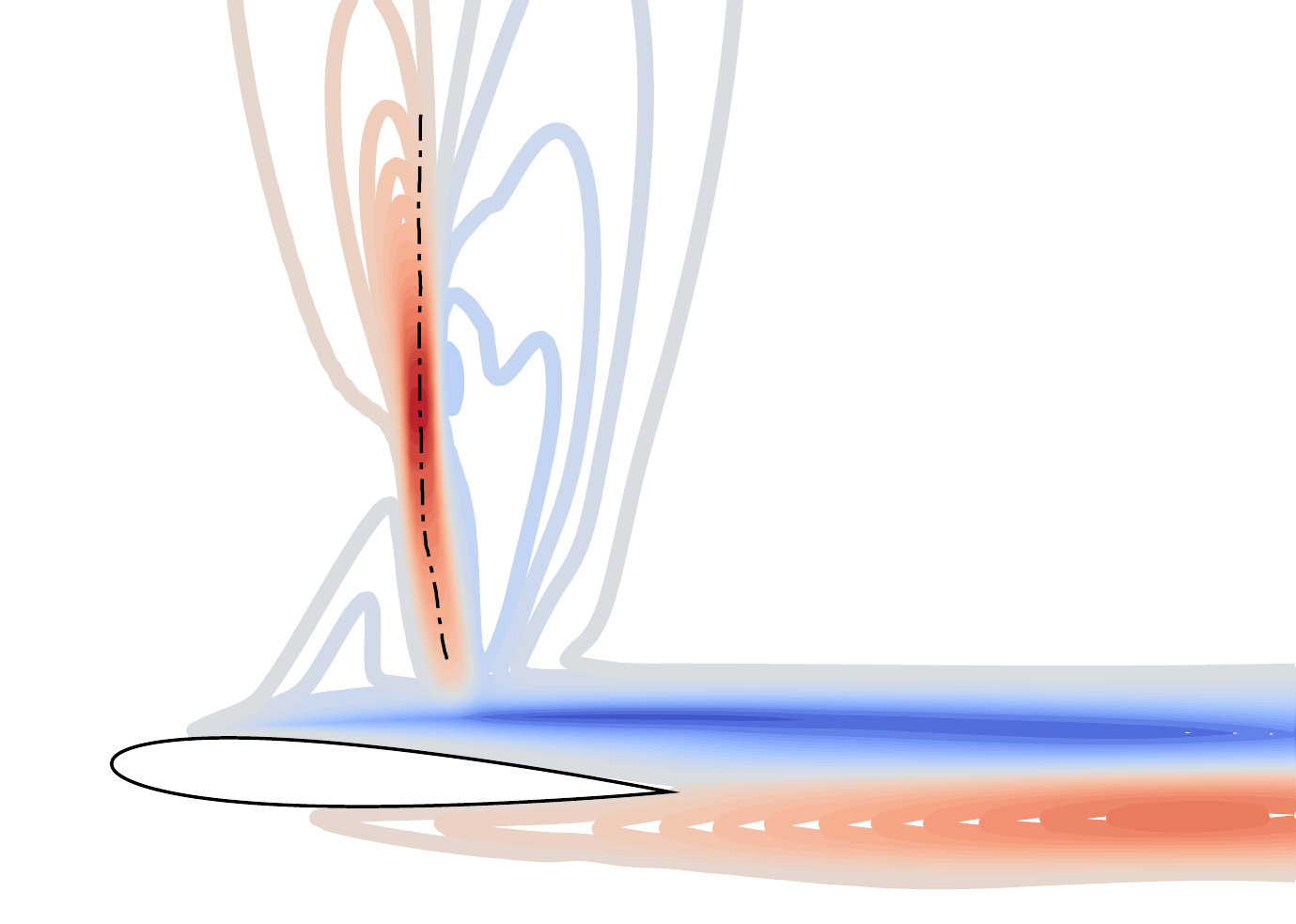} & \includegraphics[width=0.22\textwidth]{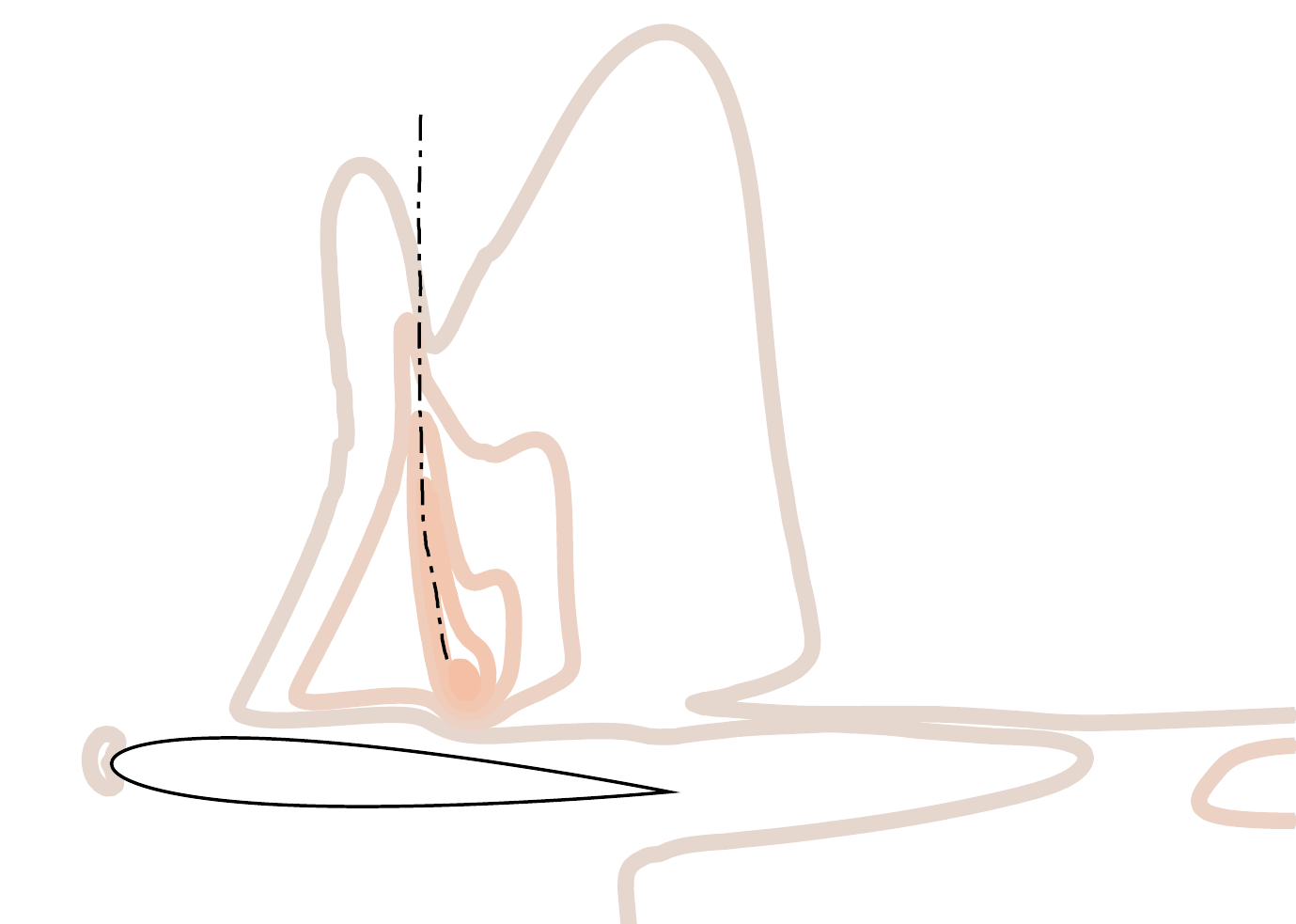} & \includegraphics[width=0.22\textwidth]{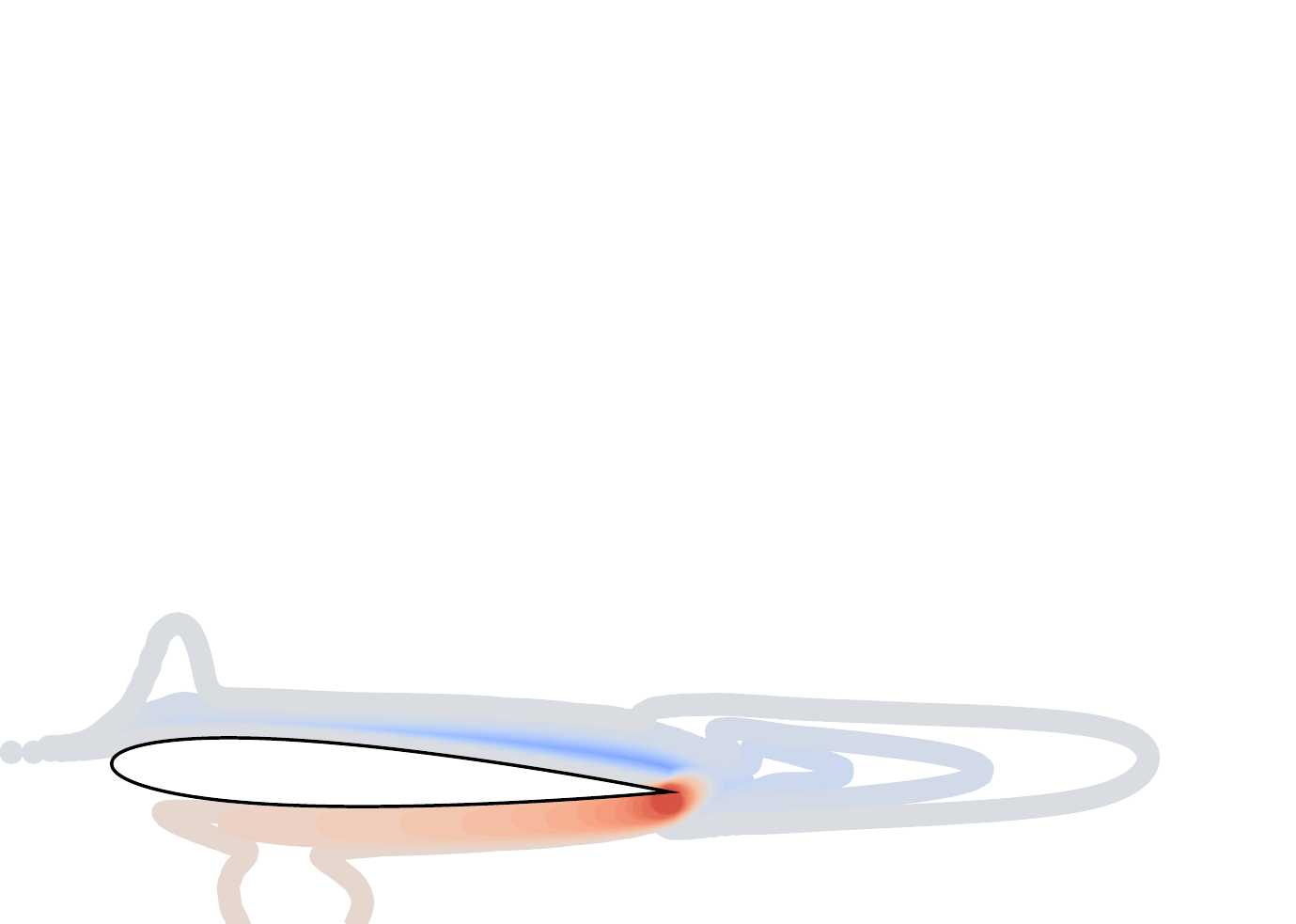} & \includegraphics[width=0.22\textwidth]{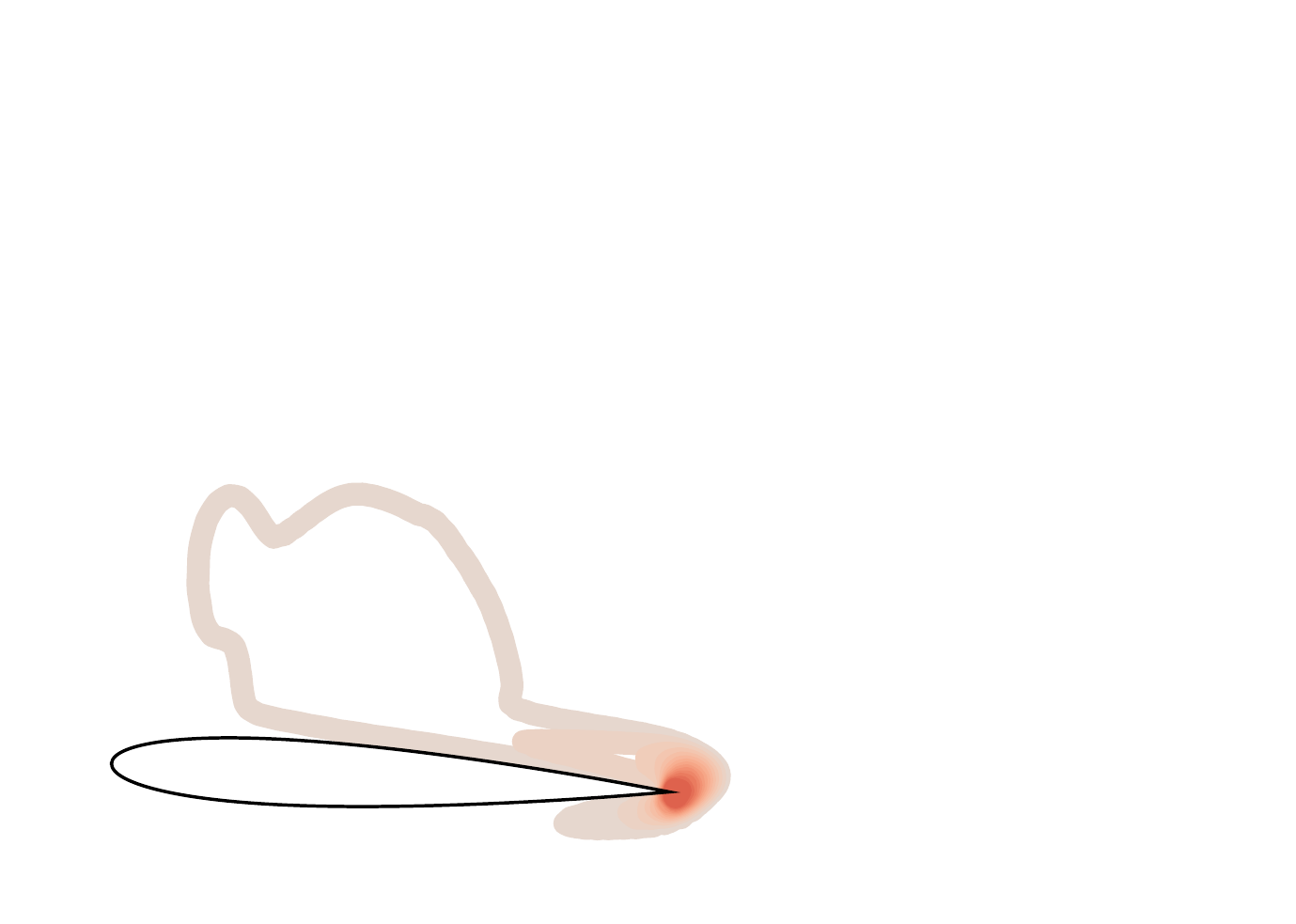} 
          \\
          \rotatebox{90}{~~Windowed} & \includegraphics[width=0.22\textwidth]{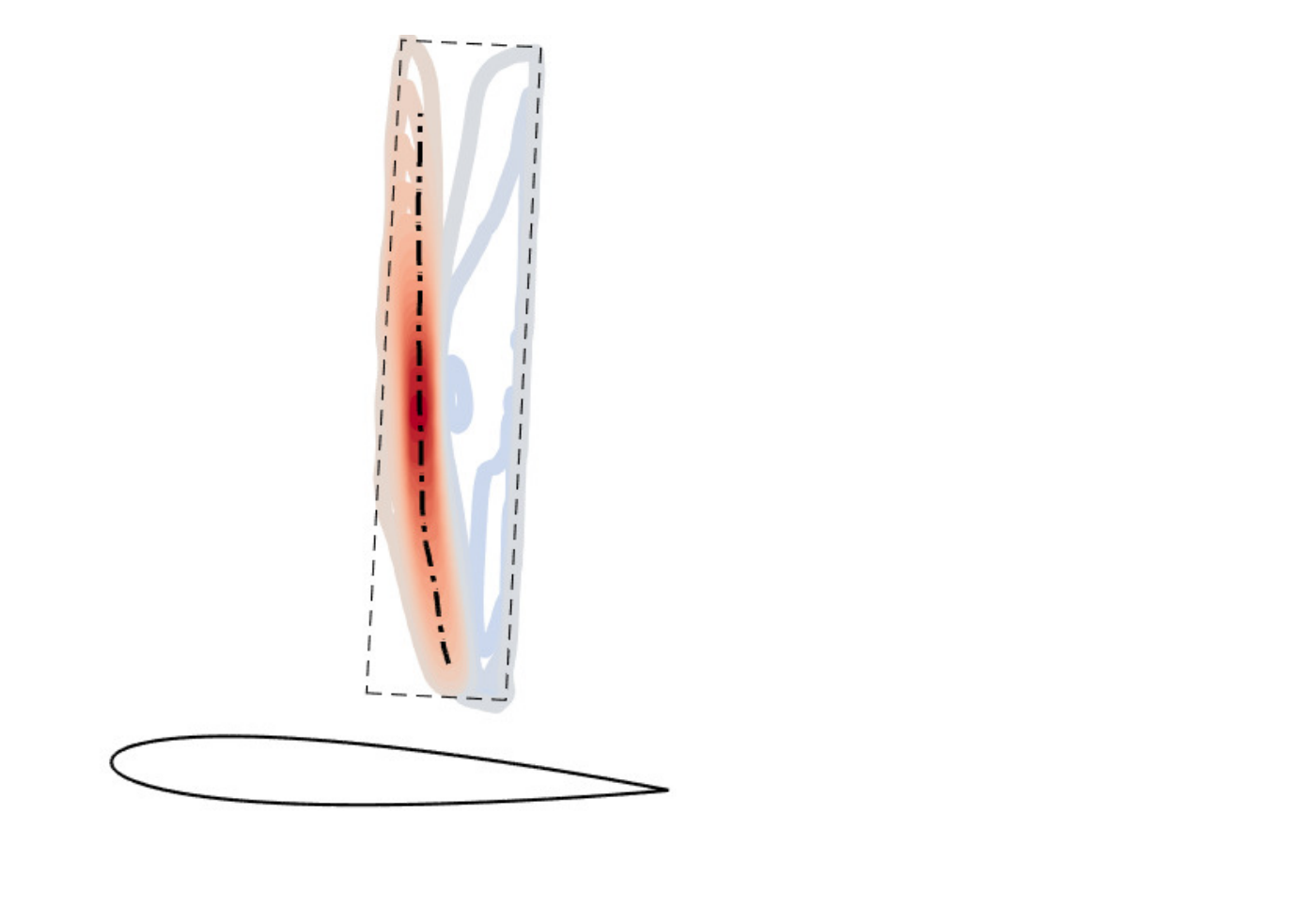} & \includegraphics[width=0.22\textwidth]{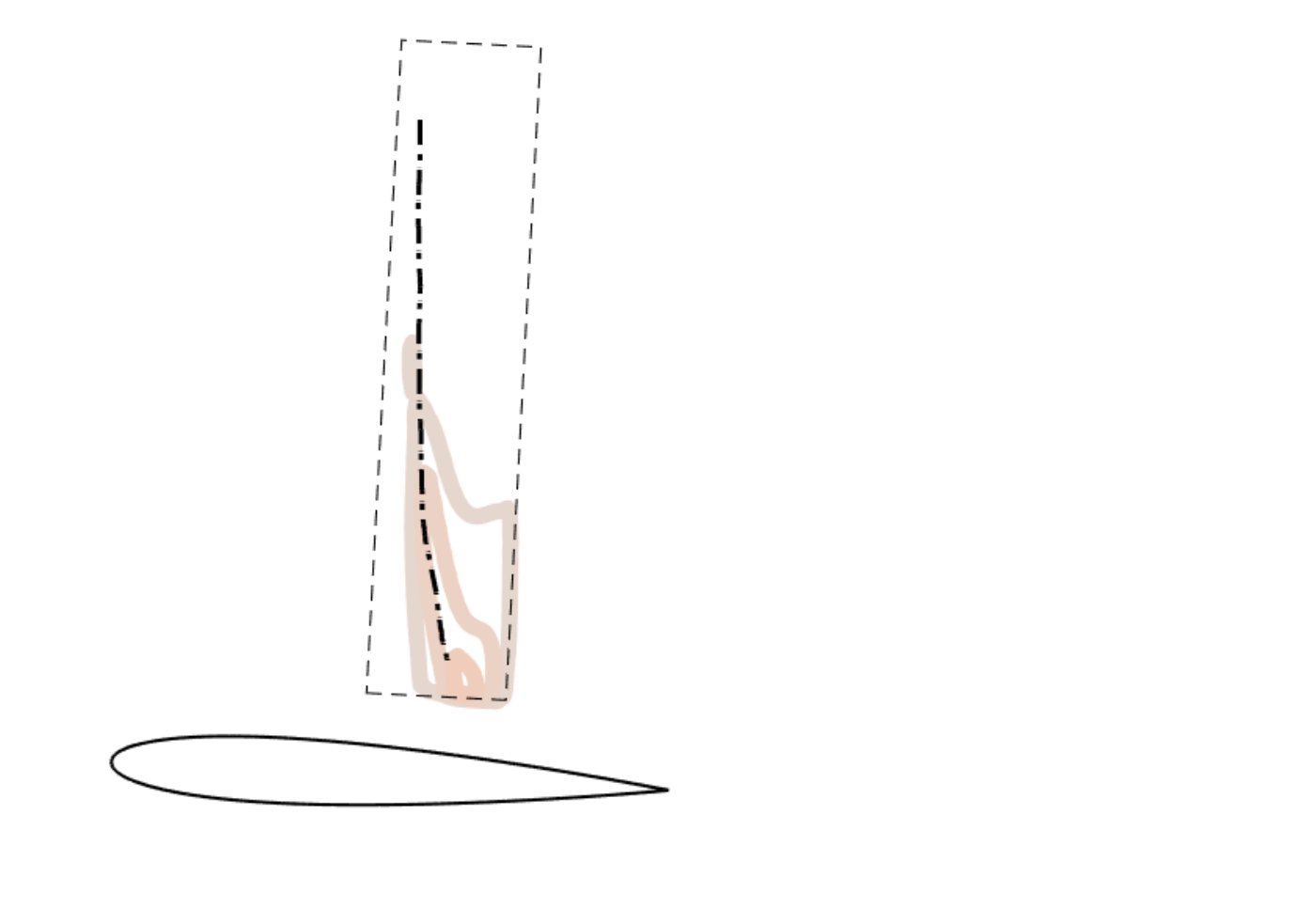} &  \includegraphics[width=0.22\textwidth]{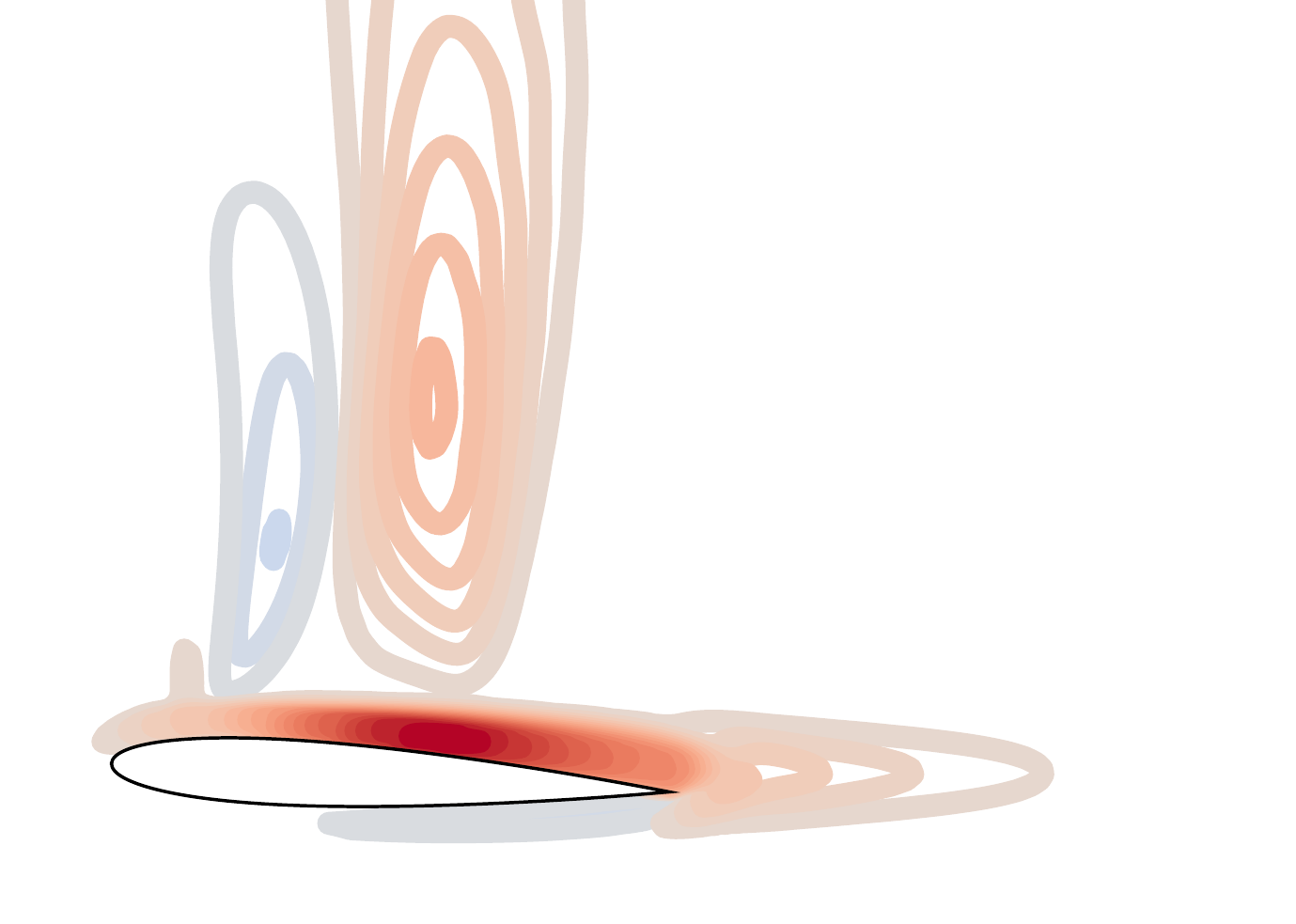} & \includegraphics[width=0.22\textwidth]{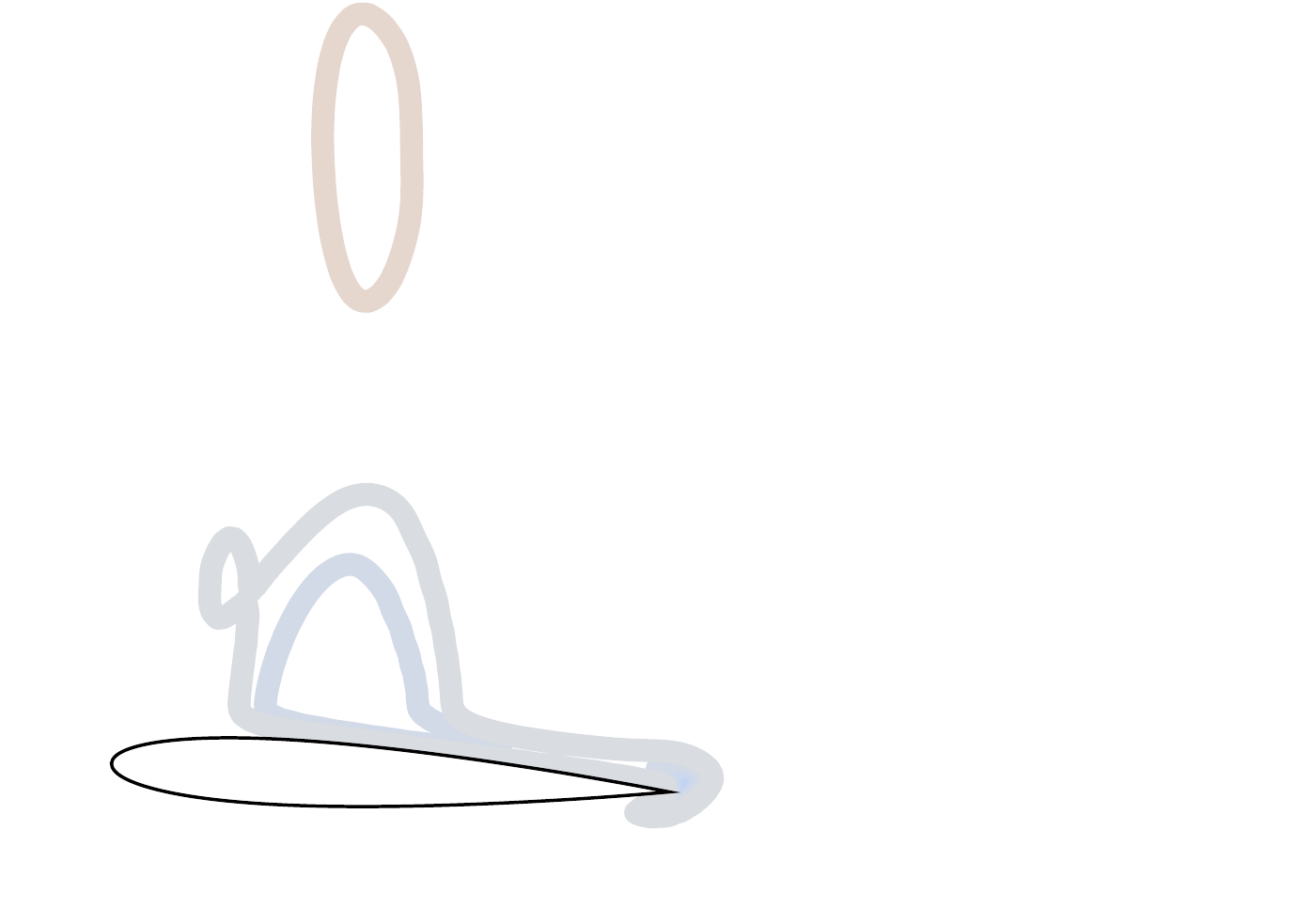}  
          \\ \hline
    \end{tabular}
    \caption{The original (top) and the windowed (bottom) resolvent modes normalized by the velocity magnitude for $St=0.06$.  The dashed-dotted lines depicts the position of the shock on the suction side.}
    \label{figMode_St006}
\end{figure}

We visualize the response and forcing modes that correspond to the buffet frequency of $St = 0.06$ as identified from the gain distribution with windowed resolvent analysis.  These modes in figure \ref{figMode_St006} are quite different from the modes at $St = 1$.  While the response modes for $St = 1$ exhibit convective oscillations in the wake region, the response modes for $St = 0.06$ are predominantly supported in the region of the standing shock.  The windowed analysis is able to highlight the response in the region of the shock.  For these reasons, the response modes can be seen as the oscillatory mechanism for buffet.

The input to generate the buffet unsteadiness can be identified by studying the forcing modes.  We can observe from figure \ref{figMode_St006} that the forcing modes from both resolvent analysis with and without windowing show the boundary layer around the airfoil to be the source of buffet.  In particular, from the windowed forcing mode, we find that the region corresponding to the shock foot is the most sensitive region to instigate the buffet phenomenon, which is in agreement with \citet{sartor2014stability} who have carried out resolvent analysis at $Re = 3 \times 10^6$.  In a number of past studies \citep{deck2005numerical, Grossi2014, fukushima2018wall}, the SWBLI at the shock foot exhibited turbulent boundary layer separation, which was suspected to play a role in the emergence of buffet. The current result by the windowed resolvent analysis suggests that the low-frequency buffet can be stimulated even though the boundary layer is entirely laminar at a significantly lower Reynolds number of $Re = 1\,000$.  Interestingly, the response modes do not show fluctuations about the weaker pressure-side shock wave observed in the numerical schlieren of figure \ref{figAveFlow}(b).

Let us also bring attention to the vicinity of the trailing edge for the original forcing modes in figure \ref{figMode_St006}.  
We can observe that there is a compact region at the trailing edge that appear in the original forcing modes.  Since the dominant response mode cover the wing surface, its fluctuation alters the circulation of the airfoil, making the trailing edge to act as sensitive location due to its singular nature from the cusp geometry.  For this reason, we see the trailing edge also supporting the fluctuations in the original forcing mode.  This observation is in tune with the report of \citet{Nitzsche2009} and past control techniques using trailing-edge deflector to effectively attenuate buffet \citep{gao2017active}.  The blockage of flow around the trailing edge by the deflector can obstruct the change in circulation in a kinematic manner.  The current resolvent analysis with windowing is able to identify these sources of buffet in a clear manner.  While the present discussion is only concerned with $\alpha = 3^\circ$, we note that similar observations are made for other angles of attack of $\alpha = 1^\circ$ and $5^\circ$.

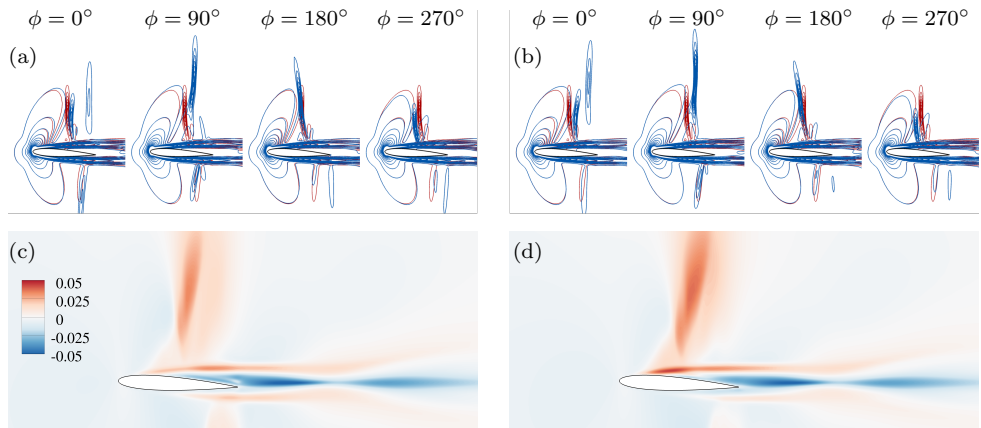
\begin{figure}
    \centering
	\vspace{0.1in}
    	\begin{overpic}[width=0.95\textwidth]{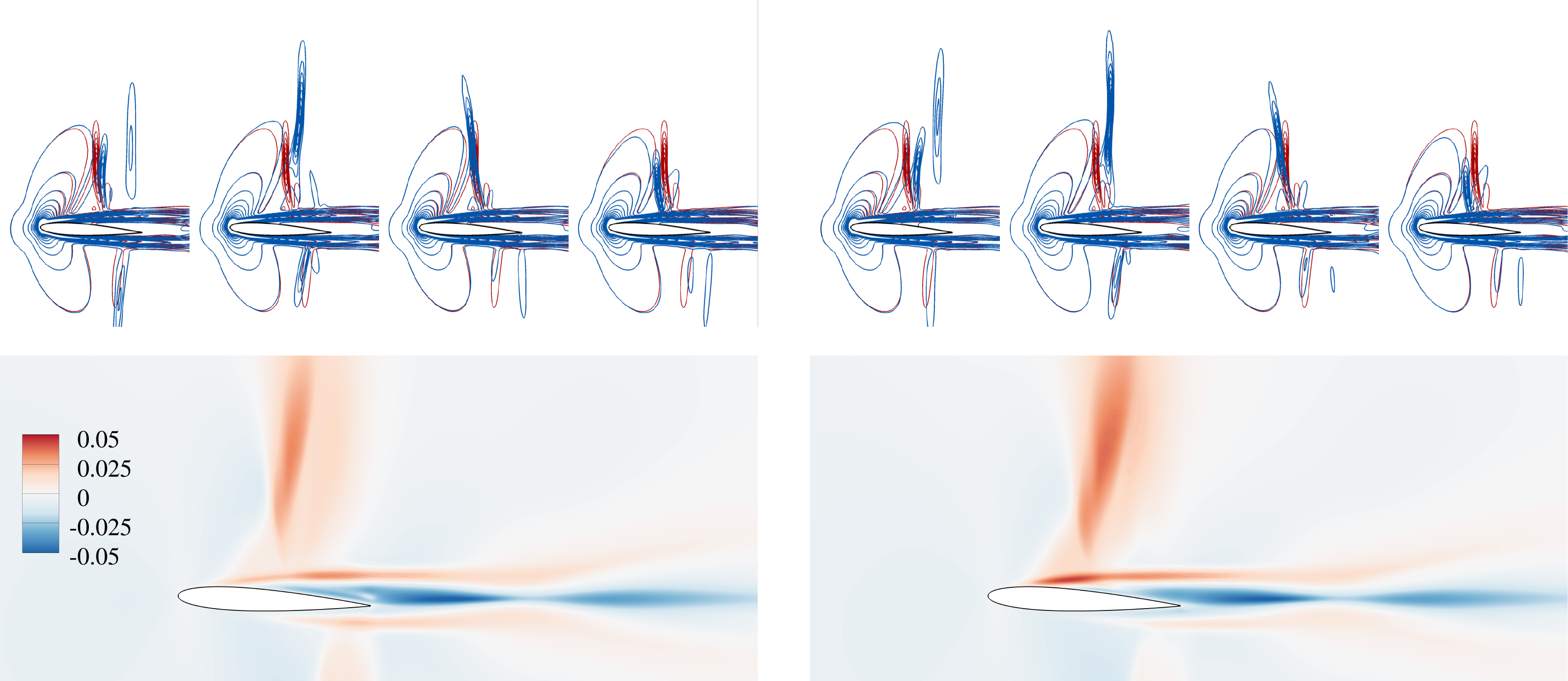}
    	\put ( 0, 38) {\indexsize (a)}
    	\put (52, 38) {\indexsize (b)}
    	\put ( 0, 18) {\indexsize (c)}
    	\put (52, 18) {\indexsize (d)}
	
    	\put ( 2, 42) {\indexsize$\phi = 0^\circ$}
    	\put (14, 42) {\indexsize$\phi = 90^\circ$}
    	\put (26, 42) {\indexsize$\phi = 180^\circ$}
    	\put (38, 42) {\indexsize$\phi = 270^\circ$}

    	\put (54, 42) {\indexsize$\phi = 0^\circ$}
    	\put (66, 42) {\indexsize$\phi = 90^\circ$}
    	\put (78, 42) {\indexsize$\phi = 180^\circ$}
    	\put (90, 42) {\indexsize$\phi = 270^\circ$}
	\end{overpic}
	\vspace{0.2in}
	\caption{
    Transonic flows over the airfoil with perturbations added in the form of the dominant forcing modes from original (left column) and windowed resolvent analysis (right column).  The motions of the shock wave are visualized in (a-b) with the blue contour lines in four phases $\phi_f$ with respect to the forced buffet cycle, on top of the red lines for unforced case.  Both contour lines display levels of $\|\nabla \rho\| L_c / \rho_\infty$. Visualized in (c-d) are differences of the rms of streamwise velocity fields $\Delta  v_{\text{rms}, x}  / a_\infty$, where $\Delta v_{\text{rms}, x} \equiv v_{\text{rms}, x, \text{forced}} - v_{\text{rms}, x, \text{unforced}}$.  
    }
    \label{figForcedFlow}
\end{figure}

To corroborate our findings from resolvent analysis, we perform companion simulations of transonic flows over the airfoil with perturbations added in the shape of forcing modes.  We add the forcing to the flow as a body force, $\boldsymbol{f}_{\text{body}}$, with the spatial profiles given by the dominant forcing modes at $St = 0.06$ (see figure \ref{figMode_St006}).  The amplitude of forcing is chosen such that the momentum coefficient
$
    C_\mu \equiv \| \boldsymbol{f}_{\text{body}} \| /(\frac{1}{2}\rho_\infty U_\infty^2 L_c) = 0.003,
$
which is low for this viscous flow problem \citep{Munday:AIAAJ18}.  These perturbations with the spatial profiles of the primary forcing modes without and with windowing do indeed stimulate the emergence of buffet over the airfoil, as shown in figure \ref{figForcedFlow} (a-b).  Even at this low Reynolds number, the standing shock is found to oscillate violently with large-amplitude motion over the airfoil.  The amplification in the oscillation from the forcing input to the flow response (including nonlinear effects) is assessed using $\Gamma = \kappa \Delta \| \boldsymbol{u}_{\text{rms}} \| / \| \boldsymbol{f}_{\text{body}} \|$, where $\Delta \| \boldsymbol{u}_{\text{rms}} \| \equiv \| \boldsymbol{u}_{\text{rms}} \|_{\text{forced}} - \| \boldsymbol{u}_{\text{rms}} \|_{\text{unforced}}$ and $\kappa \equiv \rho_\infty f^+$, showing higher amplification of $\Gamma = 9.4$ using the forcing mode from the original resolvent analysis, compared to $\Gamma = 7.0$ using that from the windowed analysis.  The higher amplification agrees with the higher gain at $St = 0.06$ obtained from the original resolvent analysis (see figure \ref{figSvDist}), which measures the global response to the forcing input.  Although the use of windowed resolvent forcing profile shows lower amplification, the shock oscillation coupled with the shear layer fluctuation at the shock foot is more energetic, as observed in figure \ref{figForcedFlow} (d).  These results validate the insights gained from resolvent analysis.  Perturbations at the foot of the standing shock becomes amplified through the linear dynamics about the base flow to oscillate the shock in a violent manner, even at a Reynolds number that is commonly not associated with buffet.

\section{Conclusions}
\label{sec:conclusion}

We examined the origin of two-dimensional transonic buffet over a NACA0012 airfoil at a Reynolds number of $2\,000$ through resolvent analysis.  
While the transonic base state at this Reynolds number exhibits unsteadiness only due to the von K\'arm\'an shedding, we show that the amplification mechanism for buffet is present in the global dynamics.  
The response and forcing modes from the windowed resolvent analysis revealed that the source of transonic buffet lies at the shock foot.  
Perturbations within the boundary layer at the shock foot can be amplified to produce oscillations about the standing shock over the airfoil with a low frequency of $St = 0.06$.  
We also noted that such perturbations are closely tied to the change in the flow around the trailing edge suggesting the effectiveness of trailing-edge buffet control devices.
The identified buffet mechanism was validated by companion DNS that generated buffet by perturbing the flow through the forcing modes.
The results show that even at a Reynolds number much lower than what is traditionally associated with transonic buffet, we are able to instigate buffet through the hidden amplification mechanisms.  
This mechanism does not require the flow to be at high Reynolds number.
The findings from the present study offers fundamental insights not only into the origin of buffet but also for low-Reynolds-number compressible aerodynamics with the growing interest in the development of Martian aircraft.

\section*{Acknowledgement}

YK and MK gratefully acknowledge the Overseas Travel Assistance Program supported by the TUAT President's Office. CAY and KT gratefully acknowledge the support from the US Office of Naval Research (Program Manager: Dr.~Brian Holm-Hansen, Grant Number: N00014-19-1-2460) and the US Air Force Office of Scientific Research (Program Managers: Drs.~Gregg Abate and Douglas Smith, Grant Number: FA9550-18-1-0040).  

\bibliography{refs}
\bibliographystyle{jfm}

\end{document}